\documentstyle[12pt,epsfig]{article}
\newskip\humongous \humongous=0pt plus 1000pt minus 1000pt
  \newif\ifdtup

\def\frac#1#2{ {{#1} \over {#2} }}
\def\sfrac#1#2{\mbox{\small $\frac{#1}{#2}$}}

\def\ltap{\raisebox{-.4ex}{\rlap{$\sim$}} \raisebox{.4ex}{$<$}}
\def\gtap{\raisebox{-.4ex}{\rlap{$\sim$}} \raisebox{.4ex}{$>$}}
\def\VEV#1{\left\langle #1\right\rangle}

\def\ie{\hbox{\rm i.e. }}

\def\beq{\begin{equation}}
\def\eeq{\end{equation}}
\def\re#1{(\ref{#1})}
\def\L{\Lambda}

\def\b{\beta}
\def\r{\rho}

\def\g{\gamma}
\def\as{\alpha_{\sf s}}

\def\bl{\beta_{\rm lat}}
\def\cl{c^{\rm lat}}
\def\cpr{c^{\rm ren}}
\def\Cpr{C^{\rm ren}}

\def\Wp{W^{\rm pert}}

\def\asl{\alpha_{\sf s}^{\rm lat}}

\def\Tr{\mbox{Tr}\;}
\def\MSbar{\overline{\rm MS}}

\def\np#1#2#3{Nucl.\ Phys.\ B#1 (19#3) #2}
\def\pl#1#2#3{Phys.\ Lett.\ #1B (19#3) #2}
\def\pr#1#2#3{Phys.\ Rev.\ D #1 (19#3) #2}
\def\prep#1#2#3{Phys.\ Rep.\ #1 (19#3) #2}
\def\prl#1#2#3{Phys.\ Rev.\ Lett.\ #1 (19#3) #2}

\begin{document}
\begin{titlepage}
\begin{flushright}
     UPRF 97-06\\
     IFUM-570-FT\\
     LTH 395\\
     May 1997 \\
\end{flushright}
\par \vskip 10mm
\begin{center}
{\Large \bf
$\L^2$-contribution to the condensate \\
in lattice gauge theory\footnote{Research supported 
in part by MURST, Italy and by EC Programme ``Human Capital and Mobility",
contract CHRX-CT92-0051 and contract CHRX-CT93-0357.}}
\end{center}
\par \vskip 2mm
\begin{center}
G.\ Burgio$\,^a$,
F.\ Di Renzo$\,^b$,
G.\ Marchesini$\,^c$
and  E.\ Onofri$\,^a$ \\[.5 em]
$^a\,${\it Dipartimento di Fisica, Universit\`a di Parma \\
and INFN, Gruppo Collegato di Parma, Parma, Italy}\\[.5 em]
$^b\,${\it Department of Mathematical Sciences, \\
University of Liverpool, United Kingdom}\\[.5 em]
$^c\,${\it Dipartimento di Fisica, Universit\`a di Milano \\
and INFN, Sezione di Milano, Italy, \\
and LPTHE, Universit\`e Paris-Sud, Orsay, France}
\vskip 2 mm
\end{center}
\par \vskip 2mm
\begin{center} {\large \bf Abstract} \end{center}
\begin{quote}
We present some evidence that in lattice gauge theory the condensate
contains a non-perturbative contribution proportional to $\L^2$, the
square of the physical scale.  This result is based on an analysis of
the Wilson loop plaquette expectation from Monte Carlo simulations and
its perturbative expansion computed to eight loops.  The analysis is
not fully conclusive since the calculations are done on a finite
lattice and one needs an extrapolation to infinite lattice.  It has
been recently suggested that in the gluon condensate a
$\L^2$-contribution could be present coming from the large momentum
behaviour of the running coupling and not connected to operator
product expansion.
\end{quote}
\end{titlepage}

\section{Introduction}
The Wilson \cite{W} operator product expansion (OPE) has been applied
\cite{ITEP} to the study of non-perturbative contributions of physical
observables in asymptotically free theories.  In these studies an
important r\^ole is played by the gluon condensate $\VEV {\as
\Tr\;F^2}$.  This quantity depends on the regularization scheme and on
a momentum scale $Q$, a subtraction point or an ultraviolet (UV)
cutoff.  According to OPE the gluon condensate has the expansion
\beq\label{OPE1} W \;\equiv\; \frac{\VEV{\as\;\Tr F^2}}{Q^4} \;=\; W_0
+ ({\L^4}/{Q^4})\;W_4 + \cdots \,, \eeq where $\L$ is the physical
scale of the theory related to the running coupling $\as=\as(Q^2)$ at
the scale $Q$.  For large $Q$ one has $\L^2/Q^2 \sim
\exp\{-1/4\pi\b_0\as\}$, with $\b_0$ the first beta function
coefficient.  Since $\L$ has no expansion in $\as$, perturbative
contributions are present only in $W_0$, which from power counting is
quartically divergent in the UV region.  The term with the power
$\L^2/Q^2$ is absent since there are no gauge invariant operators of
dimension two.  The term $W_4 $ is the ``genuine'' vacuum expectation
of the gluon condensate of dimension four.

The fact that the perturbative expansion of $W$ in $\as$ is
non-convergent \cite{Ren}, implies that the various terms $W_0,\;
W_4,\; \cdots $ cannot be identified simply by collecting
contributions with a given power of $\L$.  Indeed one expects that
$W_0$ itself contains non-perturbative terms proportional to
$\L^4/Q^4$. These contributions, needed to regularize the
non-convergent expansion in $\as$, have a simple origin.  According to
the ITEP formulation, $W_0$ is obtained from contributions of Feynman
diagrams in which the virtual (Euclidean) momenta $k$ are in the
perturbative region, \ie $k^2 \gg \L^2$.  The presence of this lower
integration bound introduces in $W_0$ terms proportional to $\L^4/Q^4$
which should be canceled \cite{ITEP1} by similar terms present in the
coefficient of the unit operator contribution in the OPE of a physical
observable.

In this paper we address the question whether the non-perturbative 
contributions in $W_0$ are only the ones discussed above which are 
proportional to $\L^4/Q^4$.
The gluon condensate is particularly suitable for discussing
this question since no contributions proportional to $\L^2/Q^2$ 
are predicted by OPE. Then one may ask whether a non-perturbative 
term of order $\L^2/Q^2$ could be present.
We perform the analysis in lattice gauge theory in which,
due to the presence of an UV cutoff, the gluon condensate has no 
UV renormalon \cite{Ren} 

By using the results on $W$ in $SU(3)$ lattice gauge theory obtained
by Monte Carlo simulations \cite{MC} and by higher order perturbative
calculations \cite{DMO} we present some evidence that in the lattice
regularization $W$ contains terms proportional to $\L^2/Q^2$.  The
analysis is not fully conclusive since the calculations are done on
lattices of finite size and one needs an extrapolation to infinite
lattice.

If the presence of terms of order $\L^2/Q^2$ in $W$ is confirmed,
one may worry that OPE could be violated.

Recently Grunberg \cite{Gr} and Akhoury and Zakharov \cite{Zak}
pointed out that the gluon condensate could contain a $\L^2/Q^2$ 
contribution not accounted for by OPE but originating from 
 from power corrections in the high frequency part of the
running coupling.  Since the perturbative Feynman diagrams for the
gluon condensate are quartically divergent in the UV region,
power contributions in the running coupling, which are highly
subleading at large momentum, could generate a $\L^2/Q^2$ term
in $W_0$. Power corrections in the running coupling at large momentum
are naturally expected in physical schemes for the coupling definition
such as the dispersive method \cite{DMW}  and the V-scheme 
\cite{BLM}. A similar observation has been given in \cite{BBB}.

The paper is organized as follows.
In Sect.~2 we recall the expected behaviour of the perturbative 
coefficients for $W_0$, a quartically divergent quantity.
In Sect.~3 we recall lattice gauge theory results on the high order
perturbative coefficients of $W$ and recall that they are
consistent with the factorial growth predicted for a 
quartically divergent quantity.
In Sect.~4 we analyze the remainder of the expansion and show the indication
for the presence of a $\L^2/Q^2$ contribution in the gluon condensate.
We show that finite size effects should not be too important.
In Sect.~5 we discuss this result and its compatibility with OPE.

\section{Factorial growth of perturbative coefficients}

In the lattice theory all frequencies are bounded by the UV cutoff
$Q=\pi/a$ with $a$ the lattice spacing.  The condensate $W$ can be
written in the general form (we assume an infinite lattice for the
moment) 
\beq\label{Int1} W =\int^{Q^2}_0\; \frac{k^2\,dk^2}{Q^4}
\;f(k^2/\L^2) \,, 
\eeq 
where $k$ is the softest virtual momentum.
This expression is based on the fact that the associated observable
has dimension four and is renormalization group invariant \cite{ITEP1}
so that the function $f(k^2/\L^2)$, which does not depend on $Q$, for
large $Q$, can be expressed in terms of a running coupling at the
scale $k^2$.

We now consider the contribution $W_0$ which is given by the large
frequency part of this integral.
At large $k^2$ the running coupling can be approximated by the
perturbative coupling $\as(k^2)$ obtained by taking into account
the first few terms of the beta function.
For instance at two loops one finds $\as(k^2)$ given in terms of
the physical scale $\L$ by
\beq\label{AS}
\L^2 \simeq
k^2\;\left(\frac{1}{4\pi b_0\as(k^2)}\right)^{b_1/b_0^2}\;
e^{-1/4\pi b_0 \as(k^2)} \,,
\;\;\;\;\;\; b_0=11/(4\pi)^2\,, \;\;\;\;\;\; b_1=102/(4\pi)^4 \,.
\eeq For simplicity we may limit the discussion to the case in which
at large $k^2$ the function $f(k^2/\L^2)$ is proportional to the
perturbative running coupling. We then consider the contribution to
$W_0$ given by 
\beq\label{Int2} 
W_0^{\rm ren}
=C\;\int^{Q^2}_{\r\L^2}\; \frac{k^2\,dk^2}{Q^4}\; \as(k^2) \,, 
\eeq
with $\r \gg 1$ to make $\as(k^2)$ within the perturbative region.
Terms with higher power of $\as(k^2)$ will give similar contributions
(see later). $W_0^{\rm ren}$ depends on the parameter $\r$ which sets
the separation of the low and high frequency part of Eq.~\re{Int1}.
Clearly the lower bound $\r\L^2$ contributes to this function by terms
proportional to $\L^4/Q^4$.

Notice that, since the integral is quartically divergent for 
$Q^2\to\infty$, the precise form of $\as(k^2)$ at large $k^2$ is very 
important. In the following we assume that at large $k^2$ the running 
coupling $\as(k^2)$ is given by the beta function at two loops. 
We introduce the variable
\beq
z \equiv z_0\left(1-{\as}/{\as(k^2)}\right)
\,,
\;\;\;\;\;\;\;
\as=\as(Q^2) 
\,,
\;\;\;\;\;\;\;
z_0=\frac{1}{3b_0}\,,
\eeq
and by using for $\as(k^2)$ the two-loop form one finds
\beq
\frac{k^2\,dk^2}{Q^4}\;\as(k^2)
\;\sim \; dz\;e^{-\b z} \;(z_0-z)^{-1-\g}
\,,
\;\;\;\;\;\;\;\;
4\pi \as  \;=\; {6}/{\b}
\,,
\;\;\;\;\;\;\;
\g=2\frac{b_1}{b_0^2}
\,,
\eeq
where higher orders in $\as(k^2)$ have been ignored.
We have introduced the inverse coupling $\b$ for the 
comparison with the lattice theory.
The integration region in \re{Int2} is mapped into the region
$0<z<z_{0_-} = z_0\,(1-\bar\b/\b)$ with 
$6/\bar \b = 4\pi \as(\r\L^2)$, and
one finds the regularized ``renormalon'' expansion
\beq\label{Wren}
W_0^{\rm ren}  = {\cal N} \int_{0}^{z_{0_-}} dz\;e^{-\b z}
\;(z_0-z)^{-1-\g}
\;=\; \sum_{\ell=1}\;\b^{-\ell}\;\{\cpr_\ell
\;+\;{\cal O}( e^{-z_0\b} )\}
\,,
\eeq
with $e^{-z_0\b} \sim \L^4/Q^4$ and $\cpr_\ell$ the renormalon 
coefficients
\beq\label{cpr}
\cpr_\ell ={\cal N'}\; \Gamma(\ell+\g)\;z_0^{-\ell}
\,.
\eeq
In the coefficients of $\b^{-\ell}$ in Eq.~\re{Wren} 
the non-perturbative corrections of order $\L^4/Q^4$ 
are essential to make the expansion convergent. 
They are coming from the lower cutoff $\r\L^2$ for the perturbative 
region and thus they depend on the parameter $\r$.
The perturbative coefficient $\cpr_{\ell}$ grows factorially
due to an infrared (IR) renormalon of dimension four.

A similar factorial growth of the perturbative coefficients is found
if one considers the contributions from higher powers of the coupling.
Then the expression \re{Int2} gives a general form of the perturbative
factorial growth with the numerical constant ${\cal N}$ which takes
into account higher order corrections.

\section{Summary of $SU(3)$ lattice gauge theory results}
We first recall the basic element of $SU(3)$ lattice gauge theory and
then the results on the high order perturbative coefficient calculations.
The action is
\beq\label{S}
S[U] = - \frac{\bl}{6}\sum_{P} \Tr (U_P+U^{\dagger}_P)
\,,
\;\;\;\;\;\;\;\;\;
4\pi \asl  \;=\; {6}/{\bl}
\,,
\eeq
where the sum extends to all plaquettes $P$ in a hypercubic
lattice in four dimensions.
The plaquette field $U_P$ is obtained from the link variable
$U_\mu(x)=\exp \{ a\;A_\mu(x)/\sqrt \bl \}$,
given by an exponential map on $SU(3)$, where $a$ is the lattice spacing.
In the continuum limit $a\to0$, at the classical level, the
plaquette tends to the Lagrangian density
and Eq.~\re{S} tends to the Yang-Mills continuous action.
In the lattice regularization one has an ultraviolet (UV) cutoff
$Q\,=\,\pi/a$, and $\asl$ is the coupling at this scale.
If the lattice is finite one has also an infrared (IR) cutoff
$Q_0=2\pi/Ma$, with $M$ the number of lattice points in each direction.

The condensate is given by the expectation value of the elementary
plaquette
\beq\label{plaq}
W_{1\times1} \equiv 1- \sfrac{ 1}{3} \VEV{\Tr U_P}
\,.
\eeq
We consider also the condensate $W_{2\times2}$, the expectation
value of the double plaquette, that is a Wilson loop around a square
 of size $2$. We denote in general by $W$ these quantities.

 $W$ can be obtained numerically by Monte Carlo simulations. The
lattice is finite and we denote by $W(M)$ the quantity computed on a
finite lattice of size $M$.  For the finite lattices considered in the
present simulations the values of $\bl$ are taken into the range
$\bl=6-6.5$.  In this region one expects that the renormalization
group properties are satisfied: the finite lattice size is not crucial
for $\bl \; \ltap \; 6.5$; the lattice discretization artifacts are
small for $\bl \; \gtap \; 6$.

\subsection{Factorial growth of lattice perturbative coefficients}
The coefficients $\cl_\ell(M)$ of the perturbative expansion of $W(M)$ 
on a lattice of size $M$
\beq\label{pert}
\Wp(M) \;=\; \sum_{\ell\ge1} \; \cl_\ell(M) \; \bl^{-\ell}
\,,
\eeq
are known up to eight loops. The first three terms have been computed 
analytically \cite{Pisa} for an infinite lattice.
Eight terms for the expansion of both $W_{1\times1}$ and
$W_{2\times2}$ have been computed numerically in Ref.~\cite{DMO}
for a lattice with $M=8$.

In Ref.~\cite{DMO} it has been shown that the growth with $\ell$ of
the first eight coefficients is consistent with the factorial behaviour
described in the previous section for a quantity of dimension
four like $W$. In the following we recall the analysis done in 
\cite{DMO,DMO1}.

In order to confirm that the computed eight coefficients growth
factorially as the renormalon coefficients in \re{cpr} 
one has to take into account two facts:
i)  the numerical calculations are done on a finite lattice;
ii) the coupling in \re{Wren} and in (\ref{S},\ref{pert}) are not
necessarily within the same regularization scheme.

i) {\it Finite size lattice}.
The numerical coefficients $\cl_{\ell}(M)$ have been computed on a
finite lattice with $M=8$ points in each direction.
The effect of the presence of a finite volume can be estimated
by putting the IR cutoff $Q_0=2\pi/Ma$ in Eq.~\re{Int2}, 
\ie $z<z_{\rm ir} =4\ln (M/2) /\b $ in \re{Wren}.
This reduces the size of the perturbative coefficients and, for 
values of $\beta$ with $z_{\rm ir} < z_0$, makes the integral
 well defined and  the perturbative expansion convergent.
We then define
\beq\label{WrenM}
W_0^{\rm ren}(M) = 
{\cal N} \int_{0}^{{\rm min}(z_{\rm ir},\,z_{0_-} )} 
dz\;e^{-\b z} \;(z_0-z)^{-1-\g}
\;=\; \sum_{\ell \ge 1}\;\cpr_\ell(M)\;\b^{-\ell}
\,,
\eeq
where $\cpr_{\ell}(M)$ are given by incomplete Gamma functions.
For $M\to\infty$ one has $z_{\rm ir} \to \infty$ and then
$\cpr_{\ell}(M)$ tend to the infinite volume coefficients of Eq.~\re{cpr}.
The reduction of the coefficients for finite $M$ been studied in detail
in \cite{DMO1}. The conclusion is that even for a small size lattice
with $M=8$, the factorial growth up to $\ell=8$ is not tamed 
(see also later).

ii) {\it Continuous and lattice coupling}.
The two coupling of eqs.~(\ref{Wren}, \ref{pert}) are not necessarily within 
the same regularization scheme. 
Differences in the couplings do not modify the factorial growth, 
but changes subasymptotic contributions. 
In \cite{DMO} we have assumed that they are perturbatively related 
at two loop by
\beq\label{rr'}
\b=\bl-r-\frac{r'}{\bl}
\,.
\eeq
It is known that the relation between the lattice and continuum coupling
involves large perturbative corrections. For instance if $\b$ is
the $\MSbar$ coupling one finds \cite{MSbar}
the values $r=1.8545$ and $r'=1.667$.
By using \re{rr'} we can connect the coefficients in \re{Wren} with the
ones in \re{pert} by computing the new coefficients $\Cpr_\ell(r,r',M)$
\beq\label{ren1}
W_0^{\rm ren}(M) \;=\; \sum_{\ell \ge 1} \; \cpr_\ell(M) \; \b^{-\ell}
\;=\; \sum_{\ell\ge1}\;\Cpr_\ell(r,r',M) \; \bl^{-\ell}
\,.
\eeq
We find that for large $\ell$ the numerical coefficients of the single 
and double plaquette
agree with $\Cpr_\ell(r,r',M)$ for large values $r=3.1$ and $r'=2.0$.
See Fig.~1. The fact that the resulting value of $r$ is larger than the
one coming from $\MSbar$ can be simply explained by assuming that the
scale entering in the  running coupling in Eq.~\re{Int2}
is $s k^2 < k^2$, in analogy to the exact result of the sigma-model
\cite{ON} (see also \cite{DMO1}).

\begin{figure}[h]\label{Coeff}
\begin{center}
\mbox{{\epsfig{figure=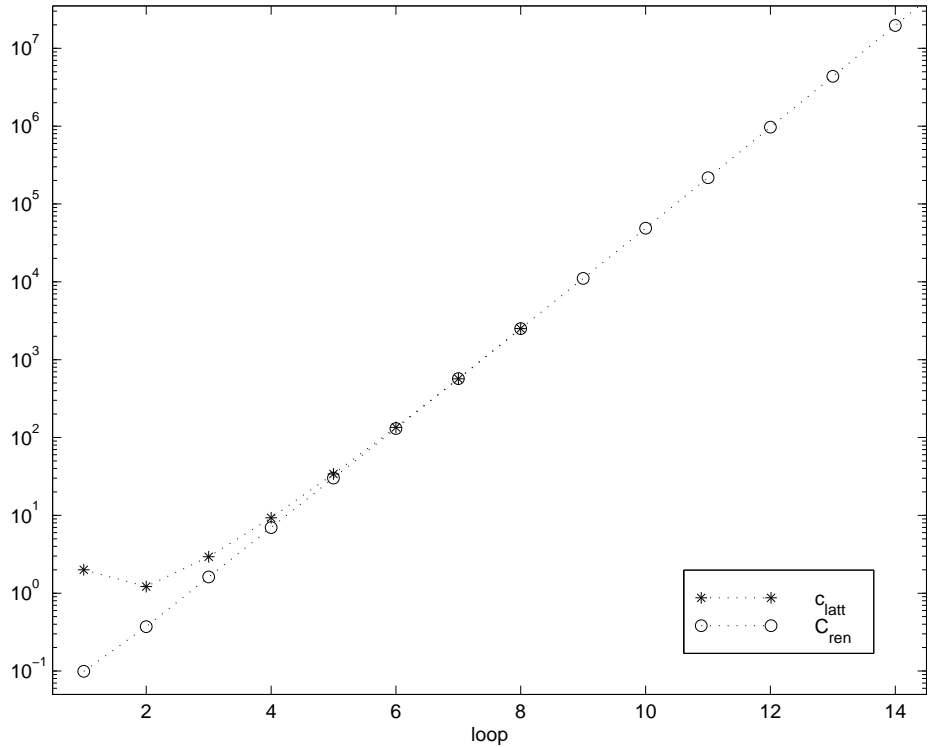,width=14.cm}}}
\end{center}
\caption{Coefficients of the perturbative expansion 
as function of the (loop) order $\ell$. 
The open squares represent $\cl_{\ell}$ from Ref.6
for $\ell \le 8$. 
The open tringles represent the renormalon coefficients 
$\Cpr_{\ell}(r,r',M)$ in Eq.~(14) for 
$r=3.1,\;r'=2.0$ and $M=8$.}
\end{figure}

Therefore the main conclusion of Ref.~\cite{DMO} was then that the
factorial growth of the computed coefficients $\cl_{\ell}$ is in
agreement with a renormalon associated to dimension four, \ie the the
gluon condensate dimension.

\section{Evidence of a $\L^2/Q^2$ contribution}

Here we present the evidence that in lattice gauge theory the
condensate contains terms of order $\L^2/Q^2 \sim e^{-z_0\b/2}$.  We
analyze $W-W_0$ as a function of $\b$.  The contribution $W$ is
obtained by the Monte Carlo simulation.  The contribution $W_0$ is
constructed by adding to the computed eight-loop perturbative terms a
remainder.  Since for large orders the perturbative coefficients of
$W_0^{\rm ren}(M)$ approach the lattice ones (see Fig.~1), we use this
function to obtain the remainder. In Eq.~\re{WrenM} we assumed that the
running coupling $\as(k^2)$ has the two-loop asymptotic behaviour.
Under this assumption the constructed $W_0$ contains, beside the
perturbative terms, power terms of order $\L^4/Q^4\sim
e^{-z_0\b}$. In this case the difference $W-W_0$ should be of order
$\L^4/Q^4\sim e^{-z_0\b}$. We shall find instead that there is a
contribution of order $\L^2/Q^2\sim e^{-z_0\b/2}$.

First we plot in Fig.~2 the quantity
\beq\label{Wsub}
\Delta_L W(M)\;=\; W(M) \;-\; \sum_{\ell=1}^{L}\cl_{\ell}(M)\;\bl^{-\ell}
\,,
\eeq
in the range $\bl=6-7$ for various values of $L \le 8$. 
The quantity $W(M)$ is obtained from the Monte Carlo
simulation of Ref.~\cite{MC} on a lattice with $M=8$ as
for the perturbative coefficients.

\begin{figure}[h]\label{fig:Wren}
\begin{center}
\mbox{{\epsfig{figure=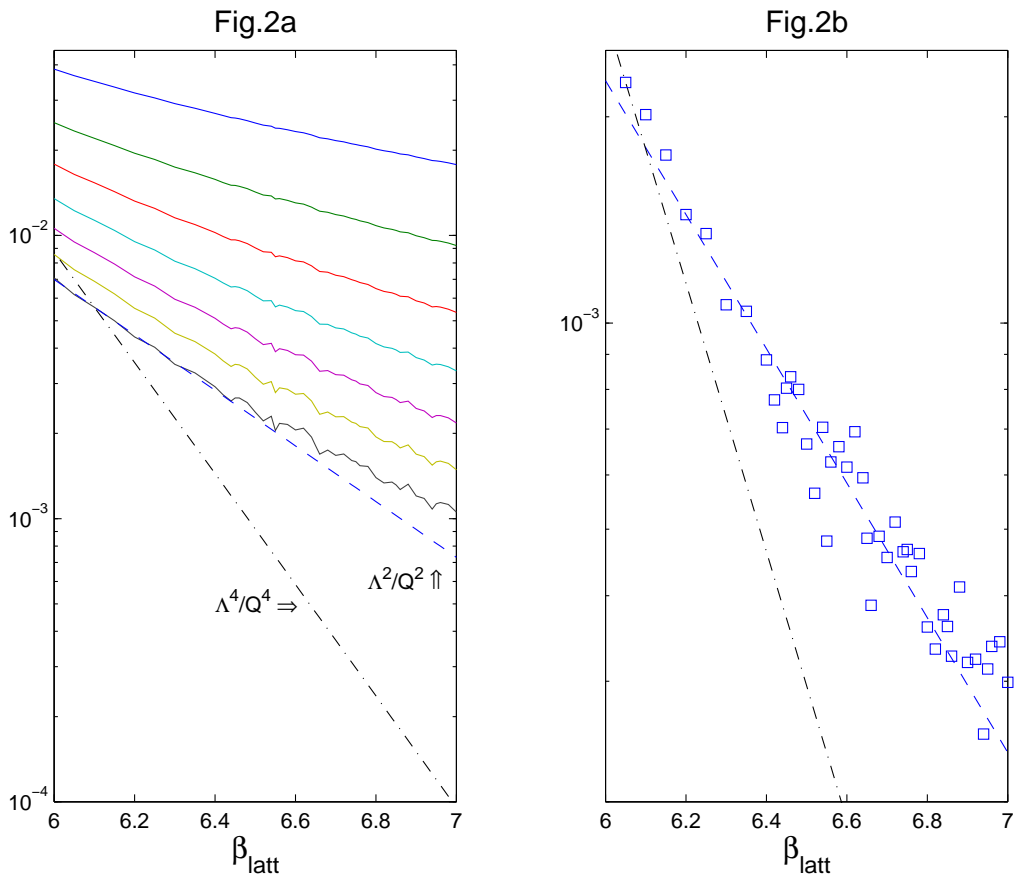,width=14.cm}}}
\end{center}
\caption{(a) The subtracted MonteCarlo data $\Delta_LW$ of Eq.~(15)
compared to $\L^2/Q^2$ and $\L^4/Q^4$ for various values of $L$: 
upper curve for $L=2$, lower curve for $L=8$; 
(b) The subtracted MonteCarlo data $\Delta W$ of Eq.~(18) 
after resummation of the renormalon contribution
compared to $\L^2/Q^2$ and $\L^4/Q^4$.}
\end{figure}

We observe that in this range of $\bl$ the quantity $\Delta_L W(M)$ approaches 
for $L\to8$ the behaviour of $\L^2/Q^2$ instead than the expected behaviour 
of $\L^4/Q^4$.

Before drawing a definite conclusion on this result one needs to
analyze whether the $\L^2/Q^2$ behaviour is modified by considering:
1) the remainder of the perturbative expansion;
2) the effects of finite volume.

\noindent
1) {\it Remainder of the perturbative expansion}.
As previously recalled, at large $\ell$ the perturbative coefficients
$\Cpr_{\ell}(r,r',M)$ in Eq.~\re{ren1}
reproduce the lattice coefficients $\cl_{\ell}(M)$.
We then estimate the remainder $\delta W_0$ by subtracting from 
$W_0^{\rm ren}(M)$, given in Eq.~\re{WrenM}, the first eight terms
in \re{ren1}
\beq\label{remainder}
\delta W_0(M) \;=\; W_0^{\rm ren}(M)-
\sum_{\ell=1}^{8}\;\Cpr_\ell(r,r',M) \;\bl^{-\ell}
\,.
\eeq
We than obtain the following estimate 
for $W_0$ 
\beq\label{W00}
W_0(M) \equiv 
\sum_{\ell=1}^{8}\cl_{\ell}(M)\;\bl^{-\ell} +\delta W_0(M) 
\,.
\eeq 
We plot in Fig.~2 the quantity
\beq
\Delta W (M) \;=\; W(M) \;-\; W_0(M)
\,,
\eeq
for $M=8$ in the region $\bl=6-7$.
We see that the behaviour $\L^2/Q^2$ is still maintained.
The conclusion is that in the region considered for $\bl$ the 
first eight perturbative terms give a reliable approximation of 
$W_0$, at least for $M=8$.

\noindent
2) {\it Finite volume}.
This effect is quite difficult to estimate without performing a direct 
Monte Carlo simulation on lattices with $M$ sufficient large to have 
the IR cutoff below the Landau singularity, \ie\  $\ln (M/2) > \b/12b_0$.
For the contribution $W_0$ we can estimate the effect of the finite 
volume. For the first eight coefficients this has been done in 
\cite{DMO1} and, as already recalled, for $M=8$ the factorial growth 
is still present. We can study the $M$ dependence of the remainder 
$\delta W_0(M)$ in \re{remainder} and we find that in the considered
region of $\bl$ the effect of finite size is small, i.e. less than 
5\% .

\section{Discussion and conclusion}

One has to consider the following two indications. 

1) As shown in \cite{DMO}, the first eight perturbative coefficients of 
$W$ seem to agree with the factorial growth corresponding to a IR
renormalon associated to an operator of dimension four, as required by 
OPE (see fig.~1).

2) Here we have studied the contribution $W_0$ in \re{OPE1}
obtained from the first eight terms of the perturbative 
expansion and a remainder \re{remainder} constructed on the hypothesis that 
only $\L^4/Q^4$ corrections are present.
By subtracting from $W$ the term $W_0$ we have found indications of 
an additional contribution proportional to $\L^2/Q^2$ (see fig.~2).

Some caveat are in order.
The reason for the unexpected behaviour $\L^2/Q^2$ could be
that our analysis is not complete. The major problem is the finiteness
of the lattice size. However we have estimated its effects on the
considered contributions to $W_0$ and they seem to be small.
It may be that the Monte Carlo simulation for $W$ contains
spurious finite size effects giving an effective $\L^2/Q^2$ behaviour.
Excluding this possibility would require an investigation on
a very large lattice with $\ln(M/2)\;\gtap\;\b$.

Recently it has been argued by Grunberg \cite{Gr} and by
Akhoury and Zakharov \cite{Zak} that terms of order $\L^2/Q^2$
can be present in the gluon condensate which are not accounted for
by OPE, but are due to power corrections in the running coupling at
high momentum. 
 In physical schemes 
\cite{DMW,BLM} highly subleading power corrections
at large momentum are naturally present in the running coupling.
A similar observation has been given in \cite{BBB}.
These corrections could be responsible for the appearance
of $\L^2/Q^2$ terms in the condensate due to the fact that
the integral for $W$ is quartically divergent. 
A $\L^2/k^2$ contribution in $\as(k^2)$ in the integral \re{Int2} 
gives two terms. The first of order $\L^2/Q^2$ is coming from the
UV region ($k^2\approx Q^2$), the second, of the canonical 
order $\L^4/Q^4$, is coming from the IR region ($k^2\approx
\r \L^2$). These $\L^2/Q^2$ terms are of ``perturbative'' nature
and are then naturally associated to the contribution $W_0$
in the OPE. Moreover they should be process independent as the running
coupling. An important question is whether these $\L^2/Q^2$ are
phenomenologically relevant (see \cite{Gr, Zak}).

\vskip 1cm
\noindent
\begin{center}
Acknowledgments 
\end{center}
\par\noindent This work is the result of discussions with
Yu.L. Dokshitzer, G. Grunberg and V.I. Zakharov.  We are most grateful
for valuable discussions with R.\ Akhoury, M.\ Beneke, V.M.\ Braun,
and A.H. Mueller.  We are indebted with L.\ Scorzato for providing us
with recent Monte Carlo data for the $SU(3)$ plaquette.


\end{document}